# Bayesian inference and superstatistics to describe long memory processes of financial time series

Geoffrey Ducournau[1]

**Abstract**

One of the standardized features of financial data is that log-returns are uncorrelated, but absolute log-returns or their squares namely the fluctuating volatility are correlated and is characterized by heavy tailed in the sense that some moment of the absolute log-returns is infinite and typically non-Gaussian [20]. And this last characteristic change accordingly to different timescales. We propose to model this long-memory phenomenon by superstatistical dynamics and provide a Bayesian Inference methodology drawing on Metropolis-Hasting random walk sampling to determine which superstatistics among inverse-Gamma and log-Normal describe the best log-returns complexity on different timescales, from high to low frequency. We show that on smaller timescales (minutes) even though the Inverse-Gamma superstatistics works the best, the log-Normal model remains very reliable and suitable to fit the absolute log-returns probability density distribution with strong capacity of describing heavy tails and power law decays. On larger timescales (daily), we show in terms of Bayes factor that the inverse-Gamma superstatistics is preferred to the log-Normal model. We also show evidence of a transition of statistics from power law decay on small timescales to exponential decay on large scale with less heavy tails meaning that on larger time scales the fluctuating volatility tend to be memoryless, consequently superstatistics becomes less relevant.

*Keywords:* Bayesian Inference, Superstatistics, Long-memory, Metropolis-Hasting sampling

## 1. Introduction

In many fields such as hydrodynamic, fluid mechanics, meteorology, traffic flows, progression of cancer cells, quantum turbulence etc. [15], there is strong evidence of existence of a phenomenon over both time and space named long memory processes. Since the late 1990's, long memory models have been playing an important role in empirical works in finance in order to emphasize substantial evidence that long memory describes well financial data characteristics such as long-term dependence and effect of shocks.

However, if this phenomenon is well known by academicians, a question raised by Grabchak and Samorodnitsky remains, do financial returns follows a Gaussian law with finite variance or an infinite variance stable law [20, 23]?

To answer this previous question, two type of time-independent descriptive models have been studying. The first one has finite-variance such as the normal distribution proposed by Osborne (1959) and Bachelier [1], the Student-t distribution proposed by Blattberg and Gonedes [6], the compound normal model proposed by Kon [19], and the mixed diffusion jump model by Merton [22]. The second one has infinite-variance symmetric and asymmetric stable Paretian distributions such as the Lévy-stable distribution proposed by Mandelbrot [2, 3, 4, 5,11] and Fama [11].

However, since the end of the twentieth century alternative solutions had emerged to answer this question by the application of econophysics concepts especially the statistical mechanics theory applies to financial problems. Particularly, numerous analogies between price dynamics and dynamics in fluid turbulence have aroused many physicians and mathematicians [25, 26].

S.M. Duarte Queiros and C. Tsallis [9, 29] are the precursor in proposing an alternative way to describe the complex behaviors of stock returns by applying the superstatistics theory, a branch of statistical mechanics developed for describing the statistics of complex system far from equilibrium characterized by large fluctuations of intensive quantities such as fluctuating financial volatility.

They have been followed by Cohen, Beck and Straeten [12, 13, 14, 15] in demonstrating that different complex system in physics and economics exhibit the same spatiotemporally inhomogeneous dynamics that can be described by a superposition of several statistics on different time scales, leading to the emergence of the terms superstatistics in Finance for the first time

[1]Geoffrey Ducournau, PhDs, Institute of Economics, University of Montpellier; G.ducournau.voisin@gmail.com

and having the advantage of relying on fundamental physical theory.

The main principle behind superstatistics rely on a strong assumption that the complex dynamics of a studied system is a superposition of two distinguishable dynamics separated by different time scales [12]: a slow dynamic related to the fluctuation of an intensive quantity and a fast dynamic related to the velocity of the system. For instance, in thermodynamics, the slow dynamic can be characterized as a slowly change in environment of the system due to a slowly fluctuating temperature $1/\theta$ ($\theta$ representing the inverse temperature) [13, 7], and the fast dynamic is determined by the change of the velocity of the studied system (such as a Brownian particle). In finance, the slow dynamic is the fluctuating volatility [12] that we will call $\theta$, and the fast dynamic is the change in velocity of the logarithmic price. And, when the slow dynamics of the fluctuating parameter $\theta$ is so slow that the velocity dynamics of the studied system (logarithm returns) has time to relax to a Gaussian distribution, after a long time, the stationary velocity distribution of the non-equilibrium statistical mechanics system (logarithm returns) becomes a superposition of infinite local statistical mechanics equilibrium characterized by the existence of respective parameter $\theta$ indexed to different time scales. If different statistical mechanics statistics can describe the variations of $\theta$, for a given system, the parameter $\theta$ dynamics will be governed by a stable law of probability with a probability density distribution called $g(\theta)$.

Indeed, it has been shown that both the log-normal superstatistics and the $\chi^2$ superstatistics were a good way to approximate the distribution of log returns [27, 28, 29]. In their article published in 2008, Biro and Rosenfeld [30] showed that the Tsallis distribution also called the q-statistics and known as being equivalent to the $\chi^2$ superstatistics is suitable to model the dynamics of the daily closing price of American indexes (Down Johns and SP500). Finally, Beck, Cohen and Swinney [8] demonstrated that the inverse $\chi^2$ superstatistics was also another superstatistics more suitable to describe daily time scales of price changes.

Theoretically, Tao Ma and R.A. Serota [31] or Gadjiev [16] show that superstatistic distribution function can be obtained from the generalized Fokker-Planck equation and proved that the Generalized Inverse Gamma distribution describes the best stock volatility dynamics and that the Students's t-distribution provides one of the better fits to returns of S&P.

In this article, we propose a Bayesian approach to determine which superstatistics among inverse gamma (IGa) and log-Normal (logN) model maximized the probability of describing the best the long memory phenomenon of financial volatility. The paper is organized as follows: in the second section we demonstrate by Bayesian Inference that the posterior probability of the fluctuating volatility $\theta$ is most likely to be proportional to an inverse gamma superstatistics. In the third section, we consider and compare two models (IGa and logN) on different time scales (minutes, hours, 4hours and daily): we first describe the empirical data used and perform timeseries analysis, we next describe the methodology and step procedures to estimate the superstatistical fluctuating parameters $\theta$ for both superstatistics on every time scales drawing on Metropolis-Hasting algorithm, we then emphasized the Bayes factor method to compare both models and determine which one is most suitable to describe fluctuating volatility respective to given timescales. Finally, the last section illustrates our conclusion on previous results.

## 2. Conjugate distribution estimate with $g(\theta)$ considered as the prior probability distribution: Bayesian inference

If we consider the previous work of Christian Beck and Eric Van der Straeten [12], it is well mentioned in their paper that in equilibrium statistical mechanics such as a conservative system, we can perfectly know how to obtain the most likely probability distributions, or the probability distributions that maximized our credence regarding the one that describe the best the long-term behavior of a system; this distribution is actually an ensemble of statistics (a canonical ensemble) governed by the famous Gibbs principle of maximum entropy [17, 21]. However, when considering financial assets' behaviors as the studied system, we are no more dealing with a conservative but a dissipative system. From a thermodynamics point of view, we know that such systems are considering as nonequilibrium systems in the same way that super statistical systems, and if their behaviors are described by a mixture of ensemble distributions, it remains very complex to be able to obtain the mixing distribution of the fluctuating parameters $\theta$.

However, as mentioned in the introduction, according to Beck and Straeten, if we consider that the dynamics of the velocity of a financial asset's log-price is enough slow to relax to a Gaussian distribution, therefore, in the long time we can consider the stationary velocity distributions of the log-price as a superposition of Gaussian distribution. Consequently, if we call the variable $x$ the observable velocity of the studied system such that for every given time $t$, $x_t$ represent the change in log-price for $t$, and we call $\mathcal{D}_n = \{x_1, ..., x_n\}$ the observing data with $t = 1, ..., n$, we are able to apply Bayes' theorem to determine the posterior probability of the fluctuating parameter $\theta$ conditioning to the data $\mathcal{D}_n$:

$$p(\theta|\mathcal{D}_n) = \frac{p(\mathcal{D}_n|\theta)\,\pi(\theta)}{\int p(\mathcal{D}_n|\theta)\,\pi(\theta)d\theta} = \frac{\mathcal{L}_n(\theta)\pi(\theta)}{c_n} \quad (1)$$

$$p(\theta|\mathcal{D}_n) \propto \mathcal{L}_n(\theta)\pi(\theta) \quad (2)$$

Where $\mathcal{L}_n(\theta) = \prod_{i=1}^{n} p(X_i|\theta)$ is the likelihood function and:

$$c_n = p(x_1, \ldots, x_n) = \int p(x_1, \ldots, x_n|\theta)\,\pi(\theta)d\theta.$$
$$= \int \mathcal{L}_n(\theta)\pi(\theta)d\theta \quad (3)$$

is the normalizing constant, which is also called the evidence.

Thus, if we use the same believe than Beck and Straeten regarding the fact that the data $\mathcal{D}_n$ is relaxing to a Gaussian distribution in the long time, and by considering that the only free parameter we have is the variance $\theta$, we know the mean $\mu$, we can consequently write the likelihood function given by Bayes' theorem:

$$p(x_1, \ldots, x_n|\theta) \propto (\theta)^{-\frac{n}{2}} \cdot \exp\left(\frac{-1}{2\theta}\sum_{i=1}^{n}(x_i - \mu)^2\right)$$
$$= (\theta)^{-\frac{n}{2}} \cdot \exp\left(\frac{-1}{2\theta} n\overline{(x - \mu)^2}\right) \quad (3)$$

Where $\overline{(x - \mu)^2} = \frac{1}{n}\sum_{i=1}^{n}(x_i - \mu)^2$.

From equation (3) we assess that the probability distribution function of the likelihood belongs to the exponential family distributions, where the sample $\mathcal{X} = \mathbb{R}_+$ is the non-negative real line and we can simplify the above likelihood distribution by:

$p(x|\theta) \sim \theta e^{-x\theta}$ (4), and in Bayesian probability, when the likelihood function is a Gaussian continuous distribution with known mean $\mu$, the conjugate prior $\pi(\theta)$ of this probability likelihood, in the most convenient parametrization, is the Inverse Gamma distribution. Given $\theta$ has an inverse Gamma distribution with parameters $\alpha$ and $\beta$ and $1/\theta \sim Gamma(\alpha, \beta)$; the density takes form as:

$\pi_{\alpha,\beta}(\theta) \propto \theta^{-(\alpha+1)}e^{-\frac{\beta}{\theta}}$ (5), and with this prior, the posterior distribution of $\theta$ is given by:

$\theta|x_1, \ldots, x_n \sim \text{InvGamma}\left(\alpha + \frac{n}{2}, \beta + \frac{n}{2}\overline{(x - \mu)^2}\right)$ (6), and is also an inverse gamma distribution function.

Moreover, the prior conjugate could also be alternatively parameterized in terms of the scaled inverse $\chi^2$ distribution with parameters $\nu_0, \sigma_0^2$, which has density of the form:

$\pi_{\nu_0,\sigma_0^2}(\theta) \propto \theta^{-\left(1+\frac{\nu_0}{2}\right)}e^{-\frac{\nu_0\sigma_0^2}{2\theta}}$ (7). Under this prior conjugate, the posterior probability distribution takes the form of:

$\theta|x_1, \ldots, x_n \sim \text{ScaledInv}-\chi^2\left(\nu_0 + n, \frac{\nu_0\sigma_0^2}{\nu_0+n} + \frac{n\overline{(x-\mu)^2}}{\nu_0+n}\right)$ (9).

Thus, from a Bayesian Inference and drawing on the assumption that the marginal likelihood of the model given some data $\mathcal{D}_n$ is a Gaussian probability distribution with known mean $\mu$ and with the variance $\theta$ as free parameter, we show that the most convenient parametrization that maximized the posterior probability is to choose a prior conjugate $g(\theta) = \pi_{\alpha,\beta}(\theta)$ that is governed by an inverse Gamma distribution of parameters $\alpha$ and $\beta$. And by demonstrating that the related distribution Scaledinv-$\chi^2$ could also be chosen as prior conjugate from a Bayesian approach, we conclude in a similar way than Beck, Xu, Straeten, Cohen, Swinney, Queiros and Tsallis that demonstrated that the inverse $\chi^2$ superstatistics is also suitable to describe the posterior probability distribution of the fluctuating parameter $\theta$.

However, even though from a Bayesian approach the inverse Gamma probability density distribution appears to be the one that maximized our credence regarding the posterior probability of the fluctuating parameters $\theta$, it is also not impossible that regarding specific time scales, related distributions to the inverse Gamma could as relevant in terms of maximizing the posterior probability. Therefore, we propose in part 3 to compare different models on different time scales with Bayesian statistics with Bayes factor testing in order to conclude which model is the most relevant for which time scales.

### 3. Model comparison and choice via Bayesian factor

#### 3.1 Empirical data

We propose in this article to use as empirical data the American Index future S&P500 drawing on the work of Lanford in 1973 [24], who argued that the level of relevance in using statistical mechanics to explain behaviors of a system such as a gas, relies on the number of particles that constitute it. Indeed, the spirit of his approach relies on the fact that the probability of errors in determining statistically the complex behaviors of a system at a macroscopic level, is inversely proportional to the number of degrees of freedom that characterized the system. More we have microscopic particles or local equilibrium dynamics with respective statistical features, more the non-equilibrium dynamics system at macroscopic scale tends to converge to a dynamics behavior relatively close to an average dynamics of every local equilibrium dynamics, describing by a superposition of statistics. For the same reason we consider index future markets as a complex and dynamics system constituting of a large number of heterogenous agents with different opportunities of profit due to large number of inhomogeneous strategies and taking decision on different time scales with different volume.

We will use one-year log returns data from 2020 corresponding to 52 weeks and propose to look at the dynamics on different time scales: minutes, hours, 4-hours and daily. We define the log returns as being the studied system, the change in log returns as being the system dynamics and we define the system fluctuation as the volatility $\theta$ and $g(\theta) = \pi_{\alpha,\beta}(\theta)$ defining the probability distribution for every given time scale. Moreover, as mentioned in part 2, we consider the system dynamics historical data $\mathcal{D}_n$ as being Normally distributed with the marginal likelihood defined by equation 4.

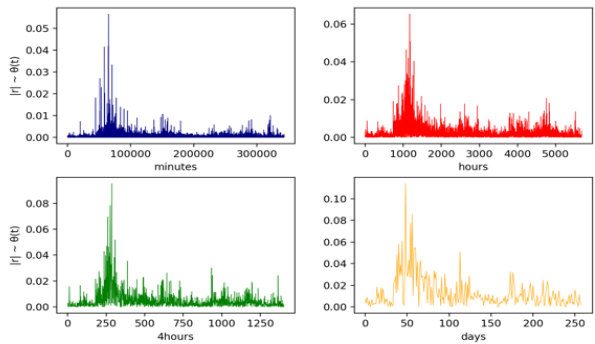

**Figure 1.** Absolute change in log - returns describes according to respective time scales (blue: $\Delta t$=minute, red: $\Delta t$=hour, green: $\Delta t$=4 hours, orange: $\Delta t$=day).

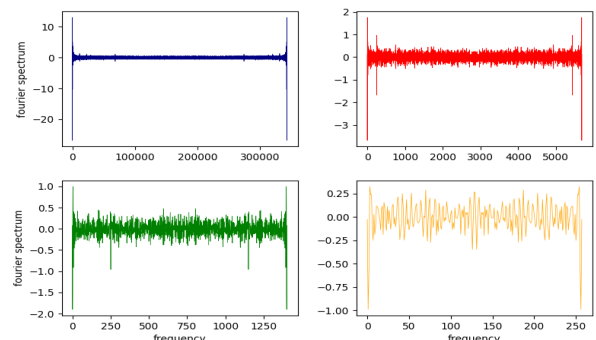

**Figure 2.** Fast Fourier transform of absolute log - returns according to respective time scales (blue: $\Delta t$=minute, red: $\Delta t$=hour, green: $\Delta t$=4 hours, orange: $\Delta t$=day).

Figure1 represents the absolute change in log-returns $|r_t| = |\log(P_{t+1}) - \log(t)|$ where $t = t_0, t_1, \ldots, t_n$ with the size of n depending on the corresponding time scales. Both the volatility clustering from time series analysis (Figure1) and the non-stationarity assessable from the spatial series analysis (Figure2) indicate a certain long-range dependence, assess confirmed by failing to reject the null hypothesis through a Dickey-Fuller test at 10% on every time scales.

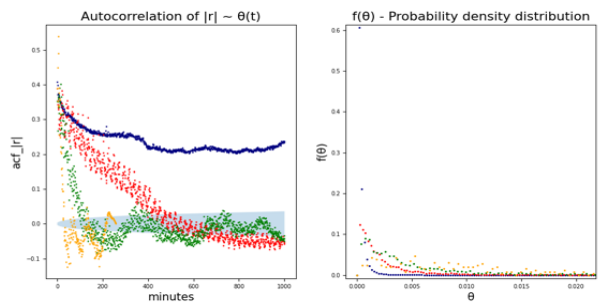

**Figure3.** figure on the left represent the autocorrelation of the financial volatility $\theta$ for the given time scales (blue: $\Delta t$ =minute, red: $\Delta t$ =hour, green: $\Delta t$ =4 hours, orange: $\Delta t$=day), figure on the right represents the probability

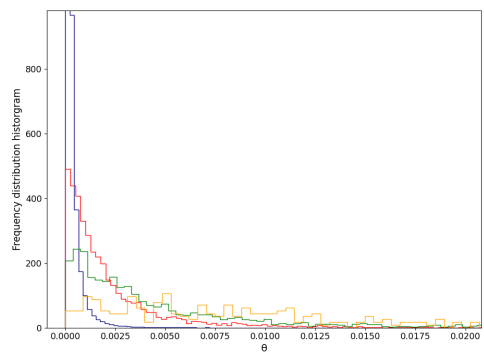

**Figure 4.** Frequency distribution diagram of $g(\theta)$ for respective time scales (blue: $\Delta t$ =minute, red: $\Delta t$ =hour, green: $\Delta t$=4 hours, orange: $\Delta t$=day

Figure3&4 gives a better overlook on long-memory phenomenon and the difference of long-range dependence according to each time scales. The autocorrelation with 5% confidence interval (Figure3) emphasized the fact that for very short-time scales $\Delta t$ = minute the autocorrelation function of the financial volatility $\theta$ is decaying more slowly than an exponential decay, and more we increase the timescales, more the decay is fast. It seems that on larger timescales such as daily of even 4-hours, the long-term dependence recedes. Figure4&5 gives us a good outlook about the existing of volatility concertation from the length of the tail distribution of $\theta$, obeying the power law distribution. But we also assess the fact that more we increase the time scale, less the distribution of the fluctuating volatility converges to a power law decay but transit towards an exponential decay with a less obvious concentration of volatility. This assessment is even more true for daily time scale where the volatility seems to be more uniformly distributed.

From these analyses' premises, we first conclude that financial volatility exhibits long term dependence as function of time scales, with a probability power law decays distribution function for small time scales with strong volatility concentration. This last sustaining the Mandelbrot multifractal model (1963) [2] to simulate volatility fluctuation using Lévy stable distribution with finite expected mean but infinite expected variance.

Second, we also conclude that on larger time scales such as 4-hours, the volatility probability density distribution and the autocorrelation function decay exponentially with remaining but smaller volatility concentration. And on very large time scales as daily, it seems that the notion of volatility concentration tends to be absent.

Next part will focus on the methodology chosen to define which model between the inverse Gamma and the log-normal superstatistics is the most relevant in terms of Bayesian criterion according to respective time scales.

### *3.2 Methodology & results on parameters estimation: Metropolis-Hasting sampler*

Before determining which superstatistics model is the most appropriate, we must priorly on every time scales estimate the parameters of each model that fit the most the observable data. By saying "fit", we mean the parameters that enable us to maximize our posterior probability of the fluctuating parameter $\theta$ conditioning to the data $\mathcal{D}_n$ as described by equation (1). To this end we introduce the Metropolis-Hasting random walk algorithm which is a specific Markov Chain Monte Carlo (MCMC) methods.

We consider equation (1) $p(\theta|\mathcal{D}_n) \propto p(x|\mathcal{D}_n) * \pi(\theta)$ as our proposal distributions, with $p(x|\mathcal{D}_n)$ the marginal likelihood and $\pi(\theta)$ the prior also called the target and being proportional to the posterior. The prior distribution is defined on a state space $\chi$ that is ergodic and stationary with respect to $\pi$, meaning that if $\theta^t \sim \pi(\theta)$ then $\theta^{t+1} \sim \pi(\theta)$ and consequently by constructing a Markov Chain on $\chi$, this last will have a unique stationary distribution that converges to the prior $\pi$. This means that the chain can be considered as a sample from the proposal distribution and the purpose is to iterate samples for given parameter $\theta$, and the acceptance criteria on $\theta$ depends on the fact that based on this parameter, the obtained stationary prior distribution $\pi(\theta)$ conjugate to the marginal likelihood must ensure the posterior distribution of interest. In other words, $\theta$ must guarantee that the prior remains proportional to the posterior.

In this paper we compare two models $\{M_1, M_2\}$:
- we call $M_1 = $ IGa as being an inverse Gamma distribution where:
$g_1(\theta; \alpha, \beta) = \frac{\beta^\alpha}{\Gamma(\alpha)}(1/\theta)^{\alpha+1}\exp(-\beta/\theta)$ (10), with $\theta > 0$ the financial volatility, $\alpha$ the shape parameter, $\beta$ the scale parameter, and $\Gamma$ is the gamma function. We define our prior distribution $\pi$ on $\theta$ conditioned to $M_1$ as $\pi(\theta|M_1) \sim g_1(\theta; \alpha, \beta)$.
- we call $M_2 = $ logN as being a log-Normal distribution where:
$g_2(\theta; s) = \frac{1}{s\theta\sqrt{2\pi}}\exp\left(-\frac{\log^2(\theta)}{2s^2}\right)$ (11), with $\theta > 0$ the financial volatility, $s > 0$ the scale parameter. We define our prior distribution $\pi$ on $\theta$ conditioned to $M_2$ by $\pi(\theta|M_2) \sim g_2(\theta; \alpha, \beta)$.

To perform the Metropolis-Hasting algorithms we proceed as following:

---

**Algorithm. Metropolis-Hasting Random Walk**

1. Generate prior $\propto$ posterior distribution:
   $\pi(\theta|M_1) \sim p(\theta|IGa; \alpha, \beta)$
   $\pi(\theta|M_2) \sim p(\theta; s|logN; s)$
2. Generate marginal likelihood distribution:
   $p(x|\theta, M_{1,2}) \sim p(x|Gaussian, \mu = 0, \theta = \sigma^2)$
3. Give two initial guess proposal $\theta_p$ and $\theta_p{'}$:
4. Calculate the ratio:
   $\rho_{M_1} = \frac{p(x|\theta_p{'}, M_1) * \pi(\theta_p{'}|M_1)}{p(x|\theta_p, M_1) * \pi(\theta_p|M_1)}$, $\rho_{M_2} = \frac{p(x|\theta_p{'}, M_2) * \pi(\theta_p{'}|M_2)}{p(x|\theta_p, M_2) * \pi(\theta_p|M_2)}$
5. Accept or reject:
   if $\rho = \{\rho_{M_1}, \rho_{M_2}\} > 1$: accept update parameter $\theta_p{'}$
   elif $\rho = \{\rho_{M_1}, \rho_{M_2}\} < 1$: reject $\theta_p{'}$ and keep $\theta_p$
6. Update parameter:
   If $\theta_p{'}$ is accepted:
      $\theta_{update} = \theta_p{'} + $ gradient $*$ learning_rate $*$ momentum
   else:
      $\theta_{update} = \theta_p + $ gradient $*$ learning_rate $*$ momentum
7. Repeat steps **4.**, **5.**, **6.**
8. End after n iterations

---

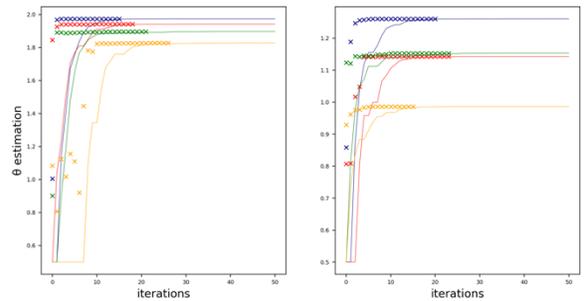

**Fig5.** Figure on the left represent the estimation of parameter $\theta$ on every time scales with the posterior probability proportional to inverse Gamma distribution (blue: $\Delta t$=minute, $\theta \sim 1.98$; red: $\Delta t$=hour, $\theta \sim 1.94$; green: $\Delta t$=4 hours, θ~1.9; orange: $\Delta t$=day, θ~1.83). Figure on the right represent the estimation of parameter θ on every time scales with the posterior proportional to log-Normal distribution (blue: $\Delta t$=minute, θ~1.26; red: $\Delta t$=hour, θ~1.14; green: $\Delta t$=4 hours, θ~1.15; orange: $\Delta t$=day, θ~0.99). The lines are the accepted updated θ and the cross are the rejected θ.

Figure5 emphasizes the Metropolis-Hasting random walk iterations for optimizing parameters θ conditioned to observed data $\mathcal{D}_n$ for both superstatistics IGa and logN. We observed a quick convergence after 25 iterations to the target parameters. If for both superstatistics parameter θ get larger on smaller time scales, we also assess a larger dispersion of θ between different time scales on the log-Normal superstatistics. Indeed, if regarding IGa superstatistics θ range from 1.83 to 1.98 from daily to minutes time scales, on log-Normal superstatistics it ranges from 0.99 to 1.26. Even though the difference is slight, it seems that the log-Normal distribution provides more information about the difference in volatility concentration between different time scales than IGa superstatistics.

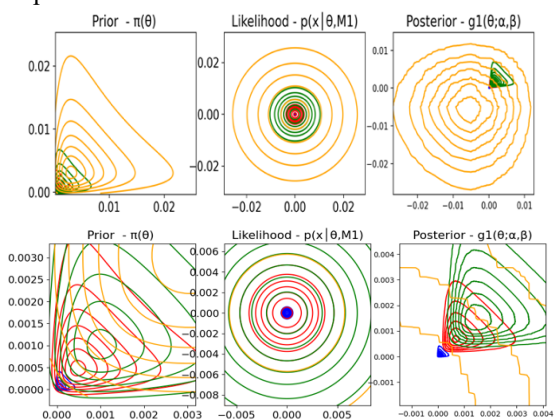

**Fig6.** Contour plot of probability density distribution on every time scales (blue: $\Delta t$=minute, red: $\Delta t$=hour, green: $\Delta t$=4 hours, orange: $\Delta t$=day) with estimated θ and inverse Gamma prior distribution.

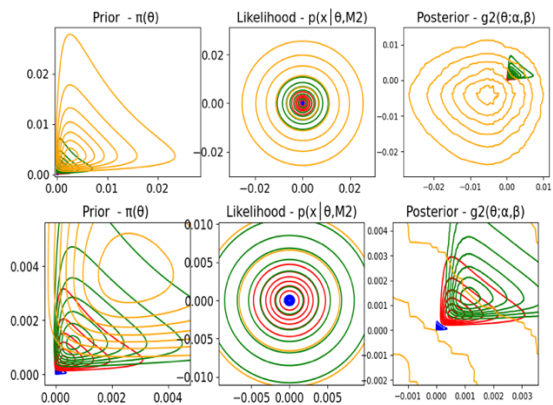

**Fig7.** Contour plot of probability density distribution on every time scales (blue: $\Delta t$=minute, red: $\Delta t$=hour, green: $\Delta t$=4 hours, orange: $\Delta t$=day) with estimated θ and log-normal prior distribution.

Figure6&7 provide another overlook of the previous assessment regarding the dispersion on every time scales of the posterior, prior and marginal likelihood probability density distribution for both superstatistics ($M_1$ for IGa & $M_2$ for log-Normal). Once again, between both model the difference in volatility concentration is slight but remains less important on model $M_2$. However, the transition of statistics from power law decay on small timescales to exponential decay on large scale with less heavy tails is easily assessable on both Figures.

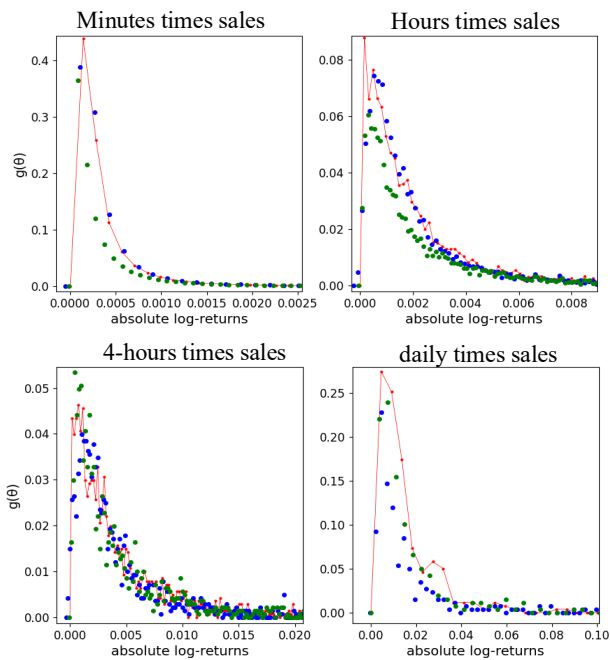

**Fig8.** Probability density distribution comparison between observed and estimated volatility on different time scales (red: observed absolute log-returns, blue: estimated absolute log-returns with IGa superstatistics, green: absolute estimated log-returns with log-N superstatistics).

Figure8 compares the probability density distribution of the observed fluctuating volatility considered as the absolute log-returns with the probability density distribution integrated from both the IGa and logN superstatistics with respective parameter $\{\theta_1, \theta_2\}$ on different time scales. If from Figure7 it was hardly possible to see a difference between both superstatistics to explain long term dependence in volatility, from Figure8 we observe clearly that the IGa superstatistics is a much better fit than the logN superstatistics to the probability distribution of empirical data by describing with accuracy the long dependence of volatility on very small-time scales with strong power lay decay. We also observe that for larger the timescales, less reliable become both superstatistics in the description of the empirical data.

Next part will consist in quantifying and comparing the degree of accuracy to fit and describe the emperical data of both the IGa and the logN model by determining the Bayes factor on every time scale through Bayesian Inference.

### 3.3 Methodology & results on models' comparison and model's choice: Bayesian factor

Once the parameter $\{\theta_1, \theta_2\}$ have been estimated for both superstatistics, we propose now to compare both model on different time scales by determining the Bayes factor between $M_1$ and $M_2$.

We consider not anymore the absolute value of the log-returns as previously but we call the series of data $R_n = r_1, \ldots, r_n$ as being the change in log-price of the index future S&P500 with the size of $n$ depending on the time scales length. We also consider two parametric model as describe previously $M_1$ and $M_2$ with $M_1$ being the IGa superstatistics with prior probability $\pi(\theta_1|M_1)$ under model $M_1$ and $M_2$ being the log-N superstatistics with prior probability $\pi(\theta_2|M_2)$ under model $M_2$.

According to bayes theorem, the posterior probability of $M_1$ and $M_2$ conditional to the data $R_n$ is:

$$\mathbb{P}(M_1, \theta_1|R_n) = \frac{p(R_n|M_1,\theta_1)\pi(\theta_1|M_1)}{\sum_{k=1}^{2} p(R_n|M_k,\theta_k)\pi(\theta_k|M_k)} \quad (12)$$

$$\mathbb{P}(M_2, \theta_2|R_n) = \frac{p(R_n|M_2,\theta_2)\pi(\theta_1|M_1)}{\sum_{k=1}^{2} p(R_n|M_k,\theta_k)\pi(\theta_k|M_k)} \quad (13),$$

where :

$p(R_n|M_1, \theta_1) = \int \mathcal{L}_1(M_1, \theta_1)\pi(\theta_1|M_1)d\theta_1$ (14)
$p(R_n|M_2, \theta_2) = \int \mathcal{L}_2(M_2, \theta_2)\pi(\theta_2|M_2)d\theta_2$ (15)

and $\{\mathcal{L}_1, \mathcal{L}_2\}$ are the likelihood function for model $M_1$ and $M_2$.
Hence the Bayes factor between both models is obtained by divided equation (14) with (15):

$$\mathcal{BF}(R_n) = \frac{\int \mathcal{L}_1(\theta_1)\pi(\theta_1)d\theta_1}{\int \mathcal{L}_2(\theta_2)\pi(\theta_2)d\theta_2} \quad (16)$$

The uses of Bayes factor are used here as a Bayesian alternative to classical hypothesis testing in the choice of models. Therefore, according to equation (16), if for a given time scales, we have $\mathcal{BF}(R_n) > 1$, then the model $M_1$ must be preferred to $M_2$. The reciprocal is also true.

Thus, we propose to draw 1000 random series of $\pi(\theta_1|M_1)$ and $\pi(\theta_2|M_2)$ in order to get 1000 Bayes factors on every time scales. The purpose here is to know the marginal probability of reliability on each model. The importance of working with many Bayes factors relies on the fact that as emphasized by Figure6&7, on very small-time scales, the difference in posterior probability density distribution of the fluctuating volatility between both superstatistics is not obvious. Therefore, determining which model in the long term gives the highest marginal probability of reliability seems more reasonable than making a decision on model based only on one iteration of bayes factor.

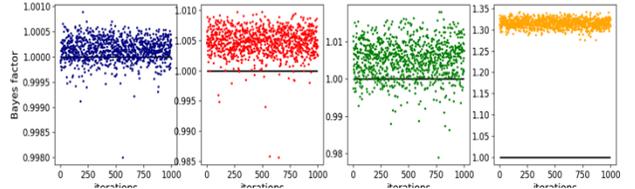

**Fig9.** Bayes factor $\mathcal{BF}(R_n)$ plotted on different time scales (blue: $\Delta t$=minute, red: $\Delta t$=hour, green: $\Delta t$=4 hours, orange: $\Delta t$=day) with the horizontal line being equal to one for 1000 iterations.

Figure9 gives a series of 1000 Bayes factor on different time scales according to equation (16) and according to estimated parameters $\{\theta_1, \theta_2\}$ respectively to model $M_1$ and $M_2$. We observe that for $\Delta t$=minute roughly 75% of the time we would prefer model $M_1$ to $M_2$ with an average Bayes factor equal to 1.00013, for $\Delta t$=hour we would prefer 98% of the time the model $M_1$ with an average Baye factor equal to 1.0048, for $\Delta t$=4 hours we would choose $M_1$ 87% of the time with an average Bayes factor equal to 1.0049, and finally for $\Delta t$=day, 100% of the time we would prefer the model $M_1$ with a Bayes factor equal to 1.31.

It seems obvious that on every time scales, the inverse Gamma superstatistics is preferable to the log-Normal, however, we also assess that more the time scales become smaller, more the superstatistics log-Normal become relevant. If we do not observe a clear transition of statistics from large timescales to small timescales as Xu and Beck [10] mentioned in their article, me must observe that on small time scales both superstatistics IGa and log-N are relatively similar with a Bayes factor slightly above to one with a precision within ten minus four.

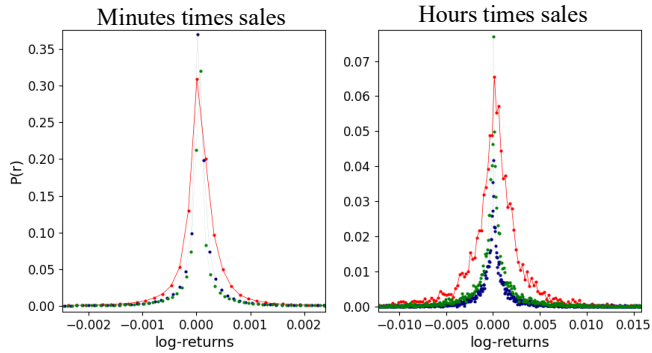

**Fig10.** Probability density distribution comparison between observed and estimated log-returns data on minutes and hours time scales (red: observed log-returns, blue: estimated log-returns with IGa superstatistics, green: estimated log-returns with log-N superstatistics).

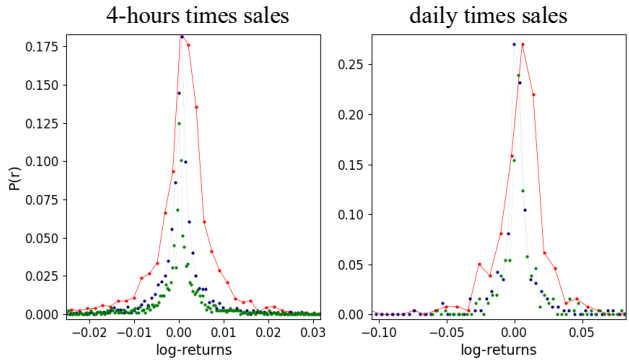

**Fig11.** Probability density distribution comparison between observed and estimated log-returns data on 4hours and daily time scales (red: observed log-returns, blue: estimated log-returns with IGa superstatistics, green: estimated log-returns with log-N superstatistics).

The Figure10&11 seems to confirm our previous conclusion obtained from the series of Bayes factor by showing that the probability density distribution of log-returns integrated from an inverse Gamma superstatistics fit better to the probability density of the observed log-returns data than the log-Normal superstatistics on every time scales. It can describe better the power law decay on shorter time scale and emphasized well the decrease in volatility concentration at longer time scale. However, we do not observe obvious transition of superstatistics from IGa to logN. If both converge to the same distributions and can describe pretty well the same distribution on shorter time scales, the superstatistics inverse Gamma remains the preferred choice on every time scales. Finally, we also observe that compare to the minute time scales, on 4hours and daily time scales, both superstatistics loose in power of accuracy in describing the probability density distribution of log-returns. The reason is that the fluctuating volatility parameter θ on shorter time scales are more correlated to each other than on longer time scale as shown in Figure3, and consequently the choice of infinite variance and volatility concentration make much less sense on daily than minutes time scales. Therefore, it would be more reasonable to describe the absolute log-returns on larger time scale with finite variance.

**Conclusion**

We have achieved satisfactory results for parameter estimation for asymmetric fluctuating volatility probability density distributions applying Bayesian inference methods, which are based on our conditionally superstatistics framework.

We first show that on small time scales particularly on minutes basis, both the power-law decay of the probability density of log-returns and the power law decline of the autocorrelations in volatilities demonstrate the presence of long dependence and strong concentration of volatility. This led us to provide evidence from Bayesian model comparison that the inverse Gamma superstatistics describe the best the fluctuating dynamics of financial volatility and must be preferred to the log-Normal model.

We first conclude that superstatistics with infinite superstatistical volatility parameter are suitable to describe this concentration volatility dependence.

Next, we provided evidence that on larger time, scales a transition of statistics takes places, with a shift from power law decay to exponential decay of the volatility probability density distribution due to the fast decline in the autocorrelations function, characteristics of memoryless phenomenon.